\documentclass{ws-p8-50x6-00}

\def\ptrel{$p_t^{rel}\,$}
\def\xgamma{$x_{\gamma}^{obs}\,$}     

\begin{document}

\title{Heavy flavor production at HERA}

\author{O. Behnke}

\address{University of Heidelberg
Philosophenweg 12
D-69120 Heidelberg, Germany\\ 
E-mail: behnke@physi.uni-heidelberg.de\\[2mm]
On behalf of the H1 and ZEUS collaborations}


\maketitle

\abstracts{
Recent results of open charm and beauty production in electron
proton scattering
at HERA are presented. 
In photoproduction, the measured cross sections 
for charm exceed fixed order NLO QCD calculations.
Experimental evidence supports the hypothesis
that a significant fraction of photoproduction events with charm  
can be described by resolved photon processes, where the
charm quark is a constituent of the resolved photon.
In deep inelastic scattering
the NLO calculations give in general a fairly 
reasonable description of the observed charm cross sections.
The measurements of the structure function $F_2^{cc}$ 
show that, at large photon virtualities and low x, 
the events with charm constitute a major part of 
the total ep cross section.
The beauty cross sections both in photoproduction and
in DIS exceed NLO predictions.
}   
\section{Introduction}
\begin{figure}[h]
\begin{center}
    \unitlength5mm
    \begin{picture}(24,6.7)
      \put(-1.,-1.7){
          \epsfig{file=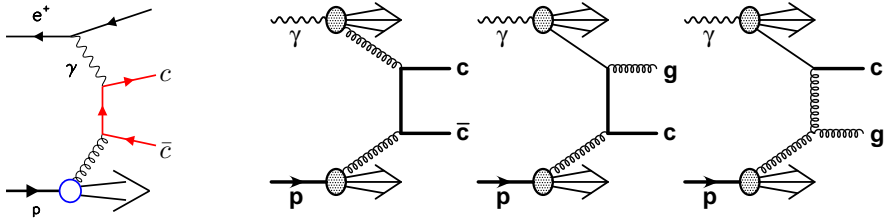,
bbllx=70.,bblly=70.,bburx=450.,bbury=400.,clip=1,width=3.5cm}}
      \put(.5, 6.4){{\bf a)} Direct $\gamma g$ fusion}
      \put(4.5,3.05){$c$}
      \put(4.5,1.0){$\bar{c}$}
\put(0.5,-9.5){\epsfig{file=hpp2.eps,
bbllx=70.,bblly=70.,bburx=450.,bbury=400.,clip=1,width=7cm}}
\put(6.,-9.5){\epsfig{file=hpp3.eps,
bbllx=70.,bblly=70.,bburx=450.,bbury=400.,clip=1,width=7cm}}
\put(11.5,-9.5){\epsfig{file=hpp4.eps,
bbllx=70.,bblly=70.,bburx=450.,bbury=400.,clip=1,width=7cm}}
      \put(11.6, 6.4){{\bf b)} Resolved $\gamma$ processes}
\end{picture}
\label{fig:bgf}
\end{center}
\caption{Leading order contributions to open heavy flavor production at HERA.}
\end{figure}

Open heavy flavor production at HERA is understood
as being mainly due to the photon gluon fusion reaction
(see Figure \ref{fig:bgf}a) ), which is directly 
sensitive to the gluon density in the proton.
In this process 
two hard scales are available facilitating the convergence of
perturbative QCD.
The first is the virtuality $Q^2$ of the exchanged
photon, with accessible values at HERA 
ranging from zero (photoproduction) to larger values 
(deep inelastic scattering, DIS), reaching 
up to $Q^2\approx 40000\,\mbox{GeV}^2$.
The second is the
mass of the heavy quark, which allows the
calculations to be extended to the photoproduction regime.
Further significant contributions 
to heavy flavor production at HERA 
are expected from {\em resolved\/}
processes (see Figure \ref{fig:bgf}b) ), in which the photon initially
fluctuates into hadronic objects
whose partons enter into the hard scattering.                                         
These processes are expected to be strongly suppressed towards larger photon
virtualities.

Within the framework of pQCD, two approaches are available to describe
heavy flavor production at HERA.
In the so-called {\em massive\/} or {\em fixed order}   
scheme\cite{massive}, 
u, d and s are the only
active flavors in the proton, and charm and beauty are
dynamically produced in the hard scattering.
It is expected that this approach works well 
at HERA for the important kinematic region  
\mbox{$p_t \le m_q$},  
where $p_t$ $(m_q)$ is the
transverse momentum (mass) of the heavy quark.
At higher transverse momenta, 
the so-called {\em massless} or {\em resummed\/}
approach\cite{massless} should be applicable, where charm and beauty
are regarded as active flavors (massless partons) in the proton 
and in the photon, 
and fragment only after the hard process into massive quarks.
This ansatz incorporates  
diagrams such as the two most right shown in Figure \ref{fig:bgf} b), which
are called charm (or beauty) excitation processes.
 
In the following we will ``take a walk'' through the
scales, i.e. in $Q^2$ from photoproduction to DIS
and in $m_q$ from charm to beauty
and compare the data for each case with QCD predictions.
The results presented here are based on analyses 
exploiting partial or full statistics of the   
HERA I data from the years 1995-2000, 
while the HERA collider was operated with 820 or 920 GeV protons
colliding on 27.5 GeV positrons or electrons.             
\section{Charm}
The results on open charm production presented here rely almost entirely
on analyses of the channel $D^{*+} \rightarrow D^0 \pi^+_s$ 
with $D^0 \rightarrow K^{-}\pi^{+}$ and the corresponding charge conjugated
mode.  

\begin{figure}[h]
\begin{center}
    \unitlength5mm
    \begin{picture}(24,9.7)
      \put(-0.25,-1.75){
          \epsfig{file=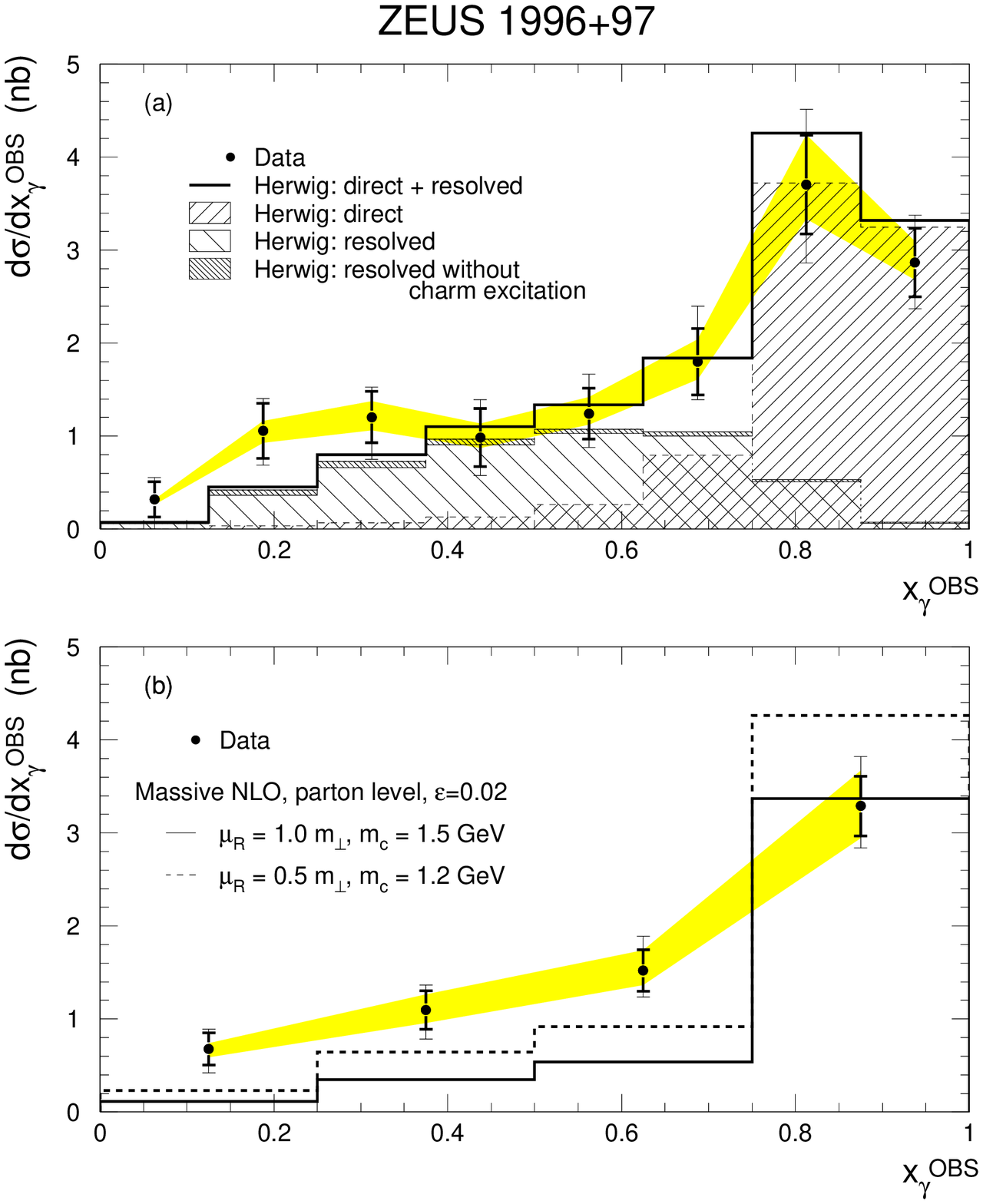,
bbllx=0.,bblly=10.,bburx=500.,bbury=450.,clip=true,width=6.cm,height=4.6cm}
}                  
      \put(12.15,-4.1){
          \epsfig{file=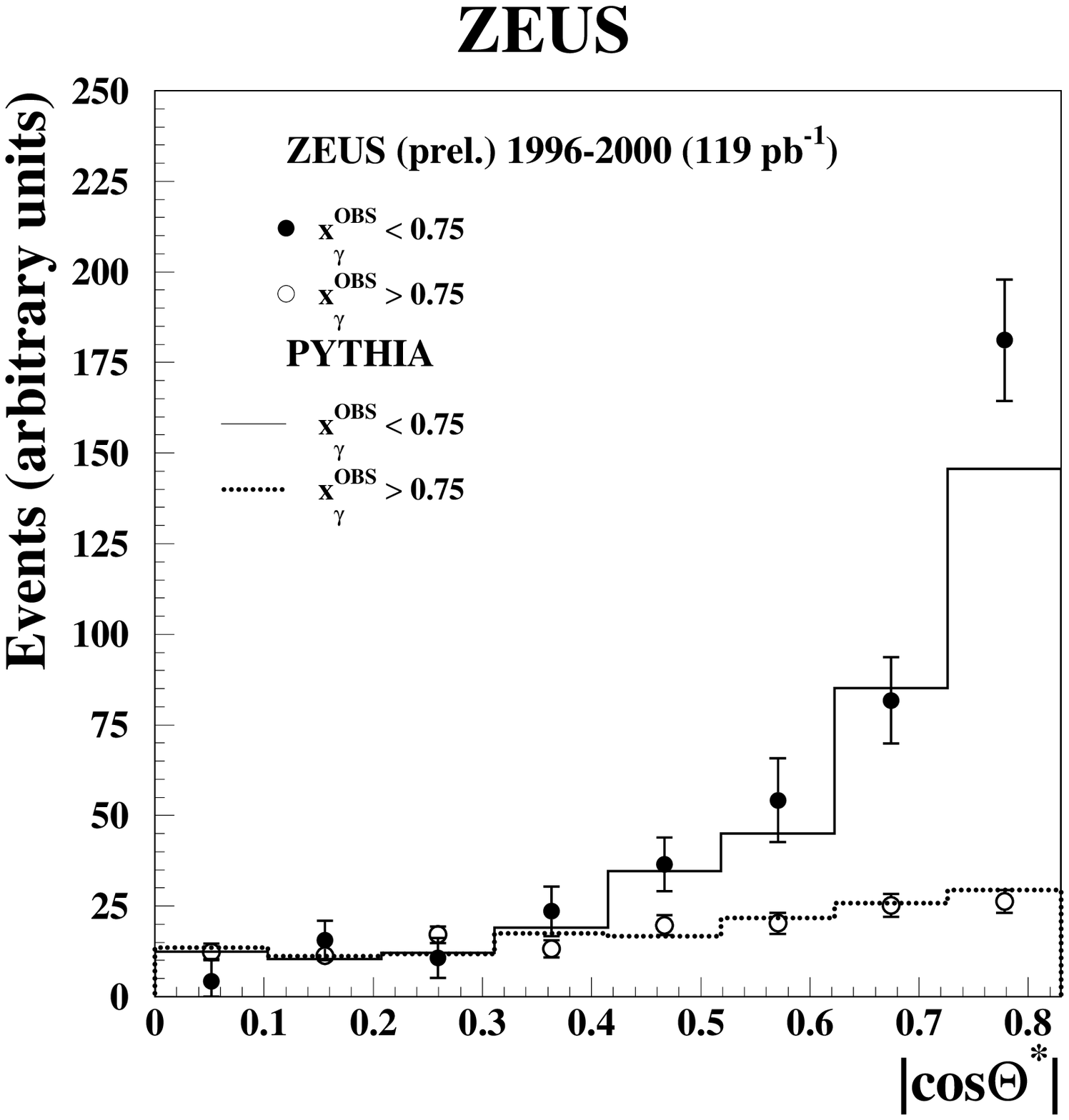,
bbllx=70.,bblly=70.,bburx=450.,bbury=850.,clip=1,width=4.6cm}}
\end{picture}                                                  
\label{fig:chphp}
\end{center}
\caption{Left: 
Differential cross section 
$\frac{d\sigma}{dx_{\gamma}^{obs}}$ in 
events with reconstructed $D^*$-mesons and two jets compared
to LO massless (top) and NLO massive (bottom) QCD
predictions. Right: Angular correlation between
the jets and the beam axis (more details in the text).
}
\end{figure}

In {\bf \em photoproduction\/}, inclusive and differential 
$D^*$-measurements\cite{zeusds} significantly exceed the massive NLO
calculations unless one uses theory parameters which are
at their extreme limits, e.g. a charm quark 
mass of 1.2 $\mbox{GeV}$. 
To obtain more information, ZEUS has measured the   
$D^{*}$-yields in events with  
two identified hard central jets\cite{zeusdstardijet1}.
From the kinematics of the jets, the photon 
fractional momentum \xgamma
entering into the hard interaction can be determined.
Figure \ref{fig:chphp} (left) shows the observed \xgamma spectrum
compared with predictions from massive NLO (bottom)
and massless LO (top) QCD.
For \xgamma values below 0.75, the domain of resolved
photon processes, the massive NLO calculation is clearly
below the data.
A much better description is obtained by the massless LO prediction, where the bulk of the resolved events is due to charm excitation processes.
In such processes, diagrams with
a gluon propagator in the hard scattering 
box (see Figure \ref{fig:bgf}b) most right) contribute.
For these diagrams, the angle between the jets and the beam
axis in the dijet center of mass system should peak more
forward than for diagrams with a quark propagator.
The angular distribution as measured by ZEUS\cite{zeusdstardijet2}
(see Figure \ref{fig:chphp} right) 
is matched well by the overlayed LO massless MC prescription
and exhibits, for the resolved part of the data (\xgamma $<0.75 $),
a stronger rise towards small angles, adding evidence to the
charm excitation hypothesis.

\begin{figure}[h]
    \unitlength5mm
    \begin{picture}(13,17.5)
      \put(1.1,-1.5){
          \epsfig{file=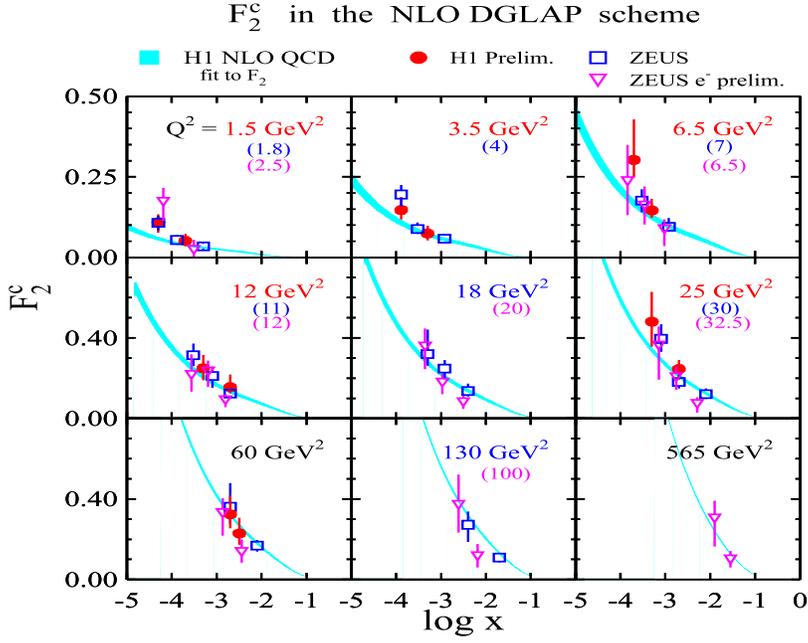,
bbllx=70.,bblly=70.,bburx=450.,bbury=850.,clip=1,width=8.cm,height=11.8cm}}
   \end{picture}
\caption{
H1 and ZEUS measurements of the structure function
$F_2^{cc}(Q^2,x)$ and comparison with 
NLO QCD theory.}
\label{f2charm}
\end{figure}

In contrast to photoproduction the
observed $D^*$-cross sections 
\cite{h1f2c}$^{,}\,$\cite{zeusf2c} 
in {\bf \em deep inelastic scattering\/}
are in general fairly well described
by the massive NLO calculations.
Some excess of reconstructed $D^*$-mesons 
is found towards the
forward region, 
which is especially sensitive to
the quark fragmentation.
A key interest in charm production in DIS
is the contribution of charm events to the
total ep scattering cross section, which can be expressed
by the structure function $F_2^{cc}(Q^2,x)$.
For this the 
$D^*$-measurements have to be extrapolated outside the accessible
$p_t(D^*)$ and $\eta(D^*)$ regions. 
The $F_2^{cc}$ measurements of H1 and ZEUS shown in 
Figure \ref{f2charm} 
agree within the errors 
with each other and with the overlayed massive NLO prediction.
The structure function $F_2^{cc}$ 
rises towards larger values of $Q^2$ and towards lower
values of x, reflecting the behavior
of the gluon density in the proton.
At highest $Q^2$ and low x, the charm contribution
to the total cross section approaches 
the expected limit of 40\%, when charm contributes
to the proton structure effectively like an u quark.
\section{Beauty}
\begin{figure}[h]
    \unitlength5mm
    \begin{picture}(13,12.)
      \put(-0.75,-0.5){
          \epsfig{file=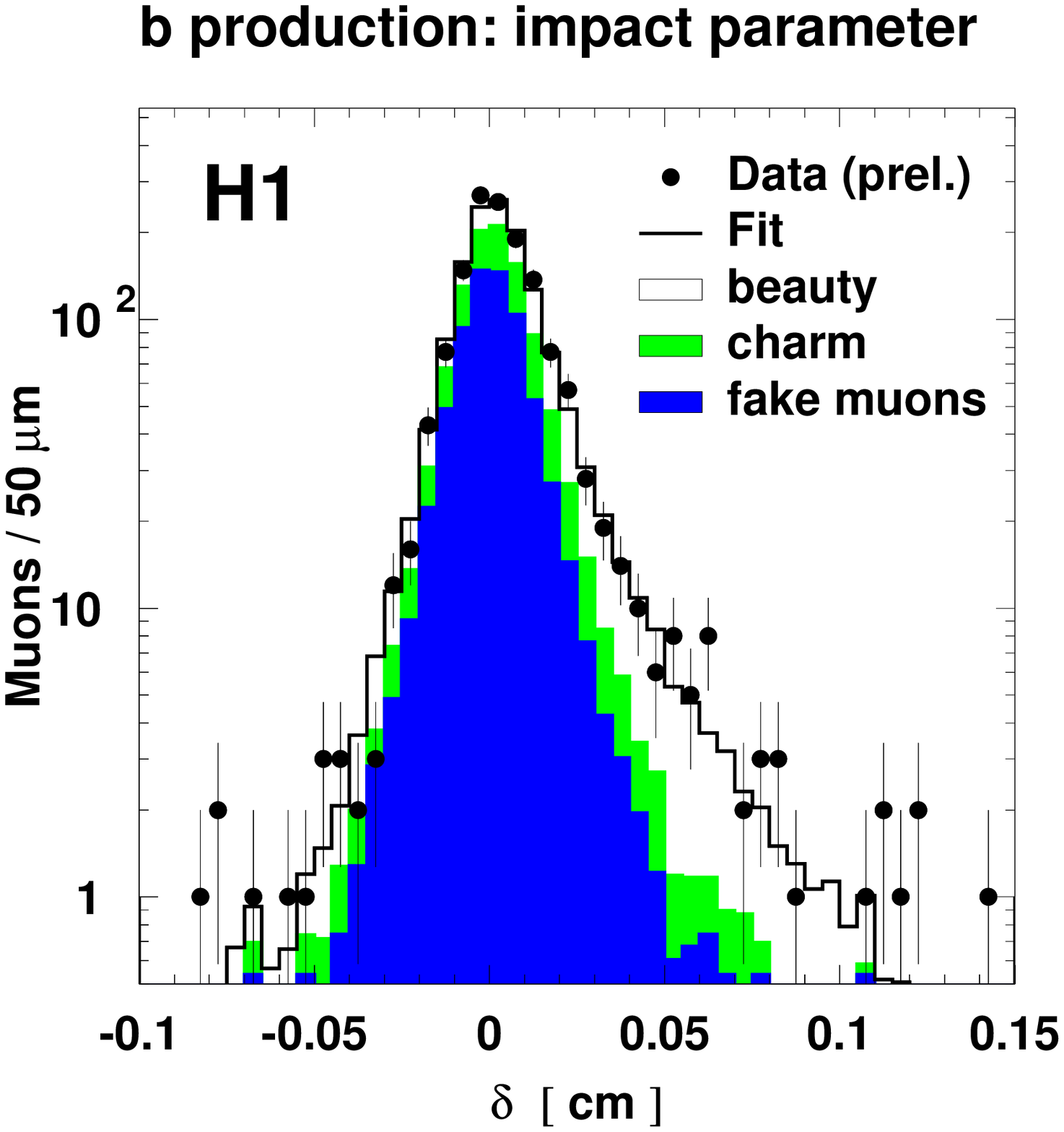,
bbllx=0.,bblly=0.,bburx=590.,bbury=840.,clip=true,width=7cm}
}
      \put(9.3,-1.4){
          \epsfig{file=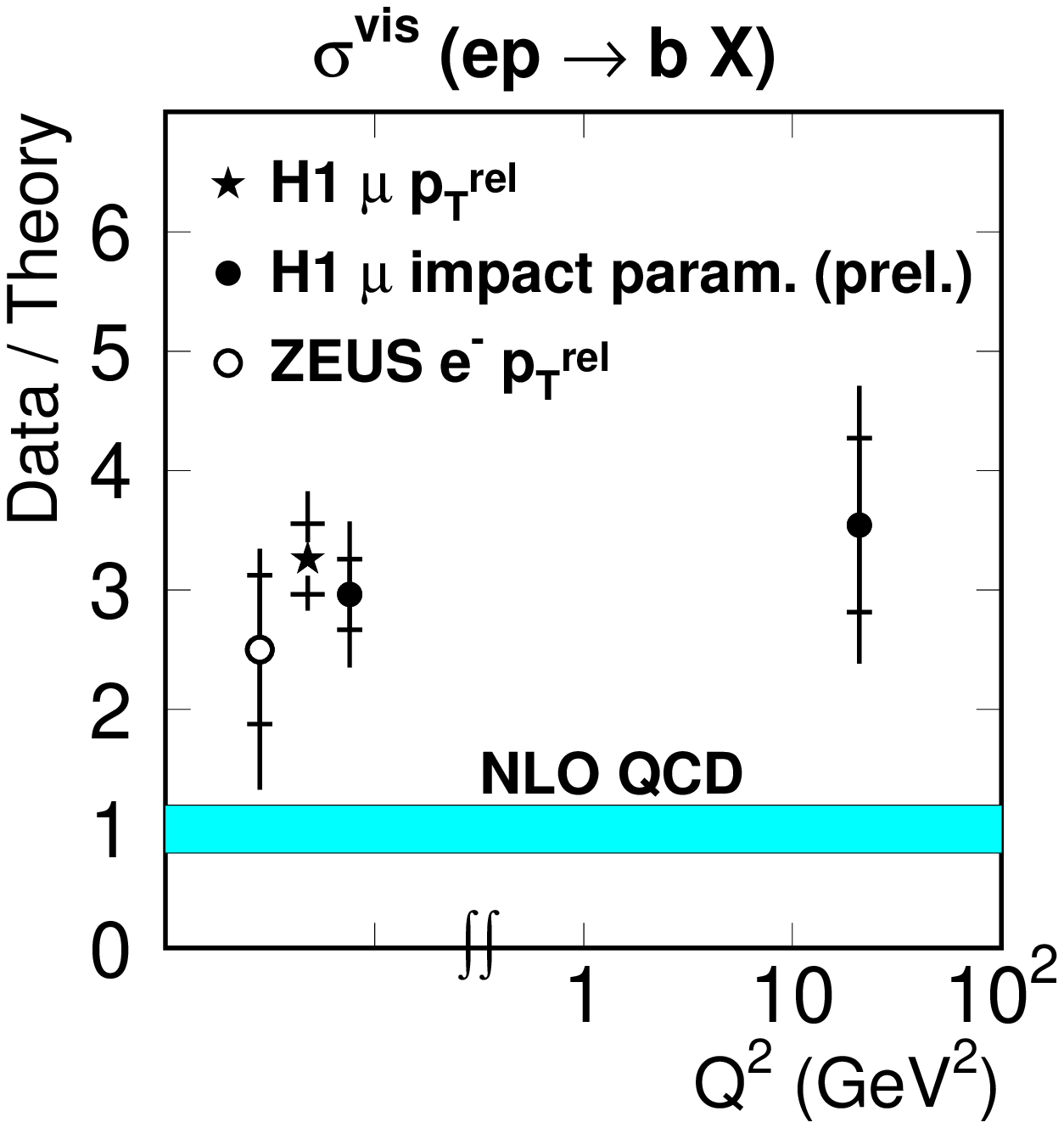,
bbllx=0.,bblly=0.,bburx=590.,bbury=840.,clip=true,width=9.85cm}
}           
\end{picture}
     \caption{
Left: H1 muon impact parameter spectrum and 
fitted signal and background contributions. Right: 
Comparison of H1 muon- and ZEUS electron-based measurements of the
b cross section with massive NLO QCD.}  
\label{fig:b}
\end{figure}

The expected beauty cross section at HERA is roughly
two orders of magnitude smaller than for charm, due
to the larger b-mass reducing the available phase space 
and the smaller b-quark charge. 
In order to find these rare processes, both H1 and ZEUS  
select dijet events with high $p_t$ leptons (muons or electrons) from 
semileptonic heavy flavor decays.
The beauty component in these events can be isolated from
charm and light quark background by using the relative
transverse momentum \ptrel of the lepton to the associated
jet, a quantity which is sensitive to the mass of the semileptonically
decaying quark.
A second observable used by H1 with its central vertex detector
is the signed impact parameter $\delta$ of muons with respect to the event
vertex. This quantity is sensitive to the quark life time.
Figure \ref{fig:b} 
shows the observed muon impact parameter spectrum in photoproduction
events together with
signal and background contributions.
The relative fractions have been determined from a likelihood fit to the 
two dimensional \ptrel versus $\delta$ distribution, yielding
a beauty component of about 27\%. 
The same measurement technique has been used by H1 to determine
for the first time the cross section for beauty production in 
the regime of deep inelastic scattering.
H1 muon and ZEUS electron results
\cite{bh1}$^{,}\,$\cite{bzeus}
have been compared
with NLO calculations in the massive scheme as shown in 
Figure \ref{fig:b} (right).
Three independent measurements in 
photoproduction and the measurement in DIS, 
with an average $Q^2$ of about 12 $\mbox{GeV}^2$ all exceed
the NLO calculations.
\section{Conclusions}
An overview of recent 
results for open charm and beauty production in ep scattering
at HERA has been presented.
In photoproduction the measured charm cross sections overshoot
the NLO QCD calculations in the massive scheme.
Experimental evidence is found 
that a significant fraction of photoproduction events with charm 
can be described by resolved photon processes, where the
charm quarks are a part of the photon structure function.
In deep inelastic scattering, the NLO QCD calculations give
a fairly reasonable description of the observed charm 
cross sections.
The measurements of the structure function $F_2^{cc}$ 
show that, at large photon virtualities and low x, the 
events with charm constitute a major part of 
the total ep cross section.
The beauty cross sections in photoproduction and in DIS, where
it is measured for the first time, significantly exceed
the NLO predictions in the massive scheme.


\begin{thebibliography}{99}
\bibitem{massive}S. Frixione {\it et al.}, \Journal{Nucl. Phys.}{B454}{3}{1995};\\
S. Frixione {\it et al.}, \Journal{Phys. Lett.}{B348}{653}{1995}.
\bibitem{massless}B.A.Kniehl {\it et al.}, \Journal{Z.Phys.}{C76}{689}{1997};\\
J. Binnewies {\it et al.}, \Journal{Z.Phys.}{C76}{677}{1997};\\
M. Cacceari and M. Greco, \Journal{Phys.Rev.}{D55}{7134}{1997}. 
\bibitem{zeusds} ZEUS Collab., J. Breitweg {\it et al.}, \Journal{Phys. Lett.}{B481}{213-227}{2000}.
\bibitem{zeusdstardijet1} ZEUS Collab., J. Breitweg {\it et al.}, 
\Journal{European Physical Journal C}{6}{67-83}{2000}.
\bibitem{zeusdstardijet2} ZEUS Collab., EPS conference 2001, Budapest, Abstract: 499. 
\bibitem{h1f2c} H1 Collab., C. Adloff {\it et al.}, hep-ex/0108039 submitted to Phys. Lett. B, 08/01.
\bibitem{zeusf2c} ZEUS Collab., J. Breitweg {\it et al.}, \Journal{European Physical Journal }{C12}{35}{2000}.
\bibitem{bh1} 
H1 Collab., C. Adloff {\it et al.}, \Journal{Phys.Lett.}{B467}{156-164}{1999};\\
H1 Collab., C. Adloff {\it et al.}, ICHEP 2000, Osaka, Abstract:979,982;\\
H1 Collab., C. Adloff {\it et al.}, EPS conference 2001, Budapest, Abstract: 807. 
\bibitem{bzeus}
Zeus Collab., J. Breitweg {\it et al.}, \Journal{European Physical Journal }{C18}{625-637}{2001}. 
%
\end{thebibliography}
\end{document}